\documentclass[a4paper,11pt]{article}
\usepackage{pos}

\usepackage[most]{tcolorbox}
\usepackage{graphicx}
\graphicspath{{figs/}}

\usepackage{mathtools}
\usepackage{slashed}
\usepackage{braket}
\usepackage{subcaption}
\usepackage[normalem]{ulem}
\usepackage{bm}

\definecolor{darkgreen}{RGB}{0,175,10}

\title{Application of $K \to 3\pi$ amplitudes to semileptonic kaon decays}

\author[a]{Anshika Bansal}
\author*[a]{Jack Jenkins}
\author[b,c]{Daniel Winney}

\affiliation[a]{Theoretische Physik 1, Center for Particle Physics Siegen (CPPS), Universit\"at Siegen, \\ Walter-Flex-Stra{\ss}e 3, D-57068 Siegen, Germany}

\affiliation[b]{Helmholtz-Institut für Strahlen-und Kernphysik (Theorie) and Bethe Center for Theoretical Physics, \\ Universität Bonn, D-53115 Bonn, Germany}

\affiliation[c]{Instituto de Ciencias Nucleares,
Universidad Nacional Aut\'onoma de M\'exico, Ciudad de M\'exico 04510, Mexico}

\emailAdd{Jack.Jenkins@uni-siegen.de}

\abstract{
We study dispersive representations of nonlocal form factors in $K^+ \to \pi^+ \ell^+ \ell^-$ and $K^+ \to \pi^+ \nu \bar{\nu}$ decays, with an aim of improving the theoretical description of the spectrum and decay rate of the neutrino mode. Based on unitarity, these representations invoke the $K^+ \pi^- \to \pi^+ \pi^-$ amplitude in $P$-wave and the pion vector form factor. The $P$-wave amplitude can be effectively parameterized within the dispersive Khuri-Treiman framework, and constrained by experimental information on the CP-conserving $K^+ \to \pi^+ \pi^+ \pi^-$ and $K^+ \to \pi^0 \pi^0 \pi^+$ decays. 
We also emphasize certain relations between charged-lepton and neutrino non-local form factors based on the Operator Product Expansion, which can be used to impose further phenomenological constraints.
}

\FullConference{13th International Conference on Kaon Physics (KAON2025)\\
8-12, September, 2025\\
Mainz, Germany\\}

\newcommand{\preprint}[1]{%
  \begin{flushright}
    #1
  \end{flushright}
}

\begin{document}

\preprint{
\begin{minipage}{3cm}
\small
\flushright
P3H-25-117 \\
SI-HEP-2025-33
\end{minipage}
}

\maketitle

\section{Introduction}
\begin{figure}[ht]
\centering
\begin{subfigure}{0.25\textwidth}
\centering
\includegraphics[width=\textwidth]{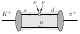}
\caption{\label{fig:figure1a}}
\end{subfigure}
\begin{subfigure}{0.25\textwidth}
\centering
\includegraphics[width=\textwidth]{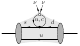}
\caption{\label{fig:figure1b}}
\end{subfigure}
\begin{subfigure}{0.25\textwidth}
\centering
\includegraphics[width=\textwidth]{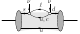}
\caption{\label{fig:figure1c}}
\end{subfigure}
\begin{subfigure}
{0.25\textwidth}
\centering
\includegraphics[width=\textwidth]{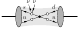}
\caption{\label{fig:figure1d}}
\end{subfigure}
\begin{subfigure}{0.25\textwidth}
\centering
\includegraphics[width=\textwidth]{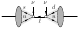}
\caption{\label{fig:figure1e}}
\end{subfigure}
\caption{Illustration of the local (a) and nonlocal (b-e) contributions to $K^+ \to \pi^+ \nu \bar{\nu}$. Only diagrams (b) and (d) contribute to $K^+ \to \pi^+ \ell^+ \ell^-$ (after replacing the local neutrino coupling with a photon propagator).}
\label{fig:figure}
\end{figure}

Precision measurements of the rare decay $K^+ \to \pi^+ \nu \bar{\nu}$ will play an indispensible role in testing the Standard Model (SM) as the theory of flavor physics at the electroweak scale. The semileptonic decay is also well-motivated, in part, because its branching ratio can be computed rather precisely; within the factorization approximation (FA), the amplitude is given by a short-distance coefficient (involving for instance the top mass~\cite{Misiak:1999yg,Buchalla:1998ba}) modulated by the standard $K \to \pi$ vector form factor (FF)~\cite{Mescia:2007kn} (see Fig.~\ref{fig:figure1a}). The largest theoretical uncertainty is therefore parametric and associated with the product of CKM matrix elements $V_{ts}^* V_{td}$, which can be extracted directly from kaon mixing ($\epsilon_K)$, or indirectly, including heavy flavor observables, from a global fit to the Unitarity Triangle.

It is well-known that, as in the case for $\epsilon_K$, contributions to $K^+ \to \pi^+ \nu \bar{\nu}$ from light flavors $q=u,c$ that are suppressed by the GIM mechanism at $O(m_c^2/M_W^2)$ are enhanced by relatively large CKM prefactors $V_{qs}^*V_{qd}/(V_{ts}^*V_{td}) = O(\lambda^{-4})$. The two mechanisms largely compensate each other, such that finite $m_c$ contributions have an $O(1)$ numerical effect. The short-distance part of these effects (roughly speaking, from the hard regions of the loop momenta $\mu \gtrsim m_c$ in Figs.~\ref{fig:figure1b}-\ref{fig:figure1c}) have been described within the FA and are under perturbative control at next-to-leading logarithmic order~\cite{Buras:2006gb}. On the other hand, effects beyond the FA (from $\mu \lesssim m_c$ and the annihilation topologies~\ref{fig:figure1d}-\ref{fig:figure1e}) may contribute at the level of several percent to the branching ratio~\cite{Falk:2000nm,Isidori:2005xm,Lunghi:2024sjy}. These `non-factorizable' effects should be understood in terms of `nonlocal FFs' defined as hadronic matrix elements (HMEs) of nonlocal operators in the appropriate low-energy effective theory renormalized at $\mu \simeq m_c$.

The recent discovery of $K^+ \to \pi^+ \nu \bar{\nu}$ and the measurement of its branching ratio by the NA62 experiment~\cite{NA62:2024pjp} motivates refinement of the description of these nonlocal FFs, at least to clarify the conceptual basis of factorization as the leading approximation. The purpose of this proceeding is to (1) introduce a correspondence between the nonlocal contributions to $K^+ \to \pi^+ \nu \bar{\nu}$ and the electromagnetic FF governing the long-distance dominated $K^+ \to \pi^+ \ell^+ \ell^-$ distributions for $\ell = e,\mu$ and (2) report on our ongoing efforts to calculate (with phenomenological input) the discontinuity of these FFs above $\pi^+ \pi^-$ threshold, following the recent developments in Refs.~\cite{Bernard:2024ioq,Bernard:2025xyn}. Regarding the first point, we will show that the FFs are anchored by dispersive representations with constraints from the Euclidean Operator Product Expansion (OPE) and isospin symmetry.

\section{Analysis of the electromagnetic form factor}

Understanding the hadronic dynamics in $K^+ \to \pi^+ \ell^+ \ell^-$ that are responsible for the observed values~\cite{NA482:2009pfe,NA62:2022qes} of the low energy constants~\cite{DAmbrosio:1998gur} would represent an important step towards reanalyzing the corresponding long distance effects in $K^+ \to \pi^+ \nu \bar{\nu}$ within a dispersive framework.  The $K^+ \to \pi^+ \ell^+ \ell^-$ distribution in the dileptonic invariant mass $q^2=(p_{\ell^+}+p_{\ell^-})^2$ is proportional to the square of the electromagnetic form factor $W_\gamma(q^2)$ defined by the following nonlocal HME
\begin{align}
    \frac{-i}{q^2}\int d^4 x \, e^{i qx}\braket{\pi^+(p)|T \{J^\mu_\gamma (x)\hat{H}(0)\}|K^+(p+q)} &= W_\gamma(q^2) \left[(2p+q)^\mu-q^\mu \frac{M_K^2-M_\pi^2}{q^2}\right] \, .\label{eq:Fgamma}
\end{align}
Here and in the following we work in QCD with $N_f=4$ quark flavors (including charm) at $\mu \sim m_c$, so the electromagnetic current is $J_\gamma^\mu = \sum_q e_q \bar{q}\gamma^\mu q$ where $e_{u,c}=2/3$ and  $e_{d,s}=-1/3$. The reduced Hamiltonian density $\hat{H}=\sum_{i=1,2}C_i(\mu)(Q_i^u-Q_i^c)$ is given in terms of the Wilson coefficients $C_{1,2}$ of the four-quark operators $Q_1^u = (\bar{s}_L \gamma_\mu T^a u_L)(\bar{u}_L \gamma^\mu T^a d_L)$, $Q_2^u = (\bar{s}_L \gamma_\mu u_L)(\bar{u}_L \gamma^\mu d_L)$ and likewise for the charm operators $Q_i^c=Q_i^u(u\to c)$. The latter enter in $\hat{H}$ with opposite sign to an excellent approximation since $V_{cs}^*V_{cd} = -V_{us}^* V_{ud} + O(\lambda^5)$. The Hamiltonian can also be written as $\hat{H}=\hat{H}_{1/2} + \hat{H}_{3/2}=C_{1/2}Q_{1/2}+C_{3/2}Q_{3/2}$ in terms of iso-tensor operators.\footnote{Fierz identities facilitating the basis change hold only in 4 dimensions. In practice we only use the decomposition $\hat{H}=\hat{H}_{1/2}+\hat{H}_{3/2}$ and do not (need to) invoke the $\Delta I = 1/2$ rule.}

The electromagnetic FF is a complex-analytic function apart from a square-root branch cut on the real axis $q^2>4M_\pi^2$, which intersects the semileptonic region $0<q^2< (M_K-M_\pi)^2$. The next branch cut $q^2>(3M_\pi)^2$ below the unitarity cut $q^2>(M_K + M_\pi)^2$ does not intersect the semileptonic region since $M_K-M_\pi<3M_\pi$. It is remarkable that the only discontinuity in the physical phase space arises from the two-body channel ($\pi \pi$). The situation here is therefore much simpler than in the heavy quark decays $D^+ \to \pi^+ \ell^+ \ell^-$ and $B^+ \to K^+ \ell^+ \ell^-$. 

Inserting a complete set of hadronic states in Eq.~\eqref{eq:Fgamma}, the two-pion contribution to the discontinuity takes the explicit form, together with the inelastic ($>2\pi$) contributions
\begin{align}
    \textrm{disc} \,W_\gamma(s) &= \frac{\sigma_\pi^3}{16\pi} [F_V(s)]^* \mathcal{A}_{\ell=1}^{K^+ \pi^- \to \pi^+ \pi^-}(s) + [\textrm{disc} W_\gamma(s)]_{>2\pi}\, ,\label{eq:discFgamma}
\end{align}
where $\sigma_\pi(s)=\theta(s-4M_\pi^2)\sqrt{1-4M_\pi^2/s}$ is the two-body phase space, $F_V^*$ is the complex conjugate of the pion vector FF and $\mathcal{A}_{\ell=1}$ is the $P$-wave projection of the $K^+ \pi^- \to \pi^+ \pi^-$ amplitude. The FF satisfies a dispersion relation with two subtractions at $q^2=-Q^2$,
\begin{align}
    W_\gamma(s) &= W_\gamma(-Q^2)+Q^2\frac{dW_\gamma(-Q^2)}{dQ^2} +\frac{(s+Q^2)^2}{\pi}\int_{4M_\pi^2}^\infty ds'\frac{ \textrm{disc}\, W_\gamma(s')}{(s'+Q^2)^2(s'-s)} \, . \label{eq:emFFdisp}
\end{align}
If we take $Q^2=0$, the subtractions $W^{(\prime)}(0)$  are related to the standard (ChPT model-dependent) low-energy constants $a_+$ and $b_+$. Fewer subtractions would be needed to predict one or both low energy constants, at the expense of exposing high-energy contributions to the discontinuity.

The subtractions for $Q \gg \Lambda$ can also be approximately \emph{calculated} in terms of the local $K \to \pi$ form factor, in the context of an OPE in $\Lambda/Q \ll 1$, where $Q\sim m_c$ and $m_s \sim \Lambda$. The OPE of the time-ordered product in Eq.~\eqref{eq:Fgamma} reads
\begin{align}
    &\frac{-i}{q^2}\int d^4 x \,e^{iqx} \,T \{J_\gamma^\mu(x) \hat{H}(0)\} = c_3(Q,m_c) \left(g^{\mu \nu}-\frac{q^\mu q^\nu}{q^2}\right) [\bar{s}_L\gamma_\nu d_L] + \dots  \, , \label{eq:ope}
\end{align}
where $c_3=c_3(Q, \overline{m}_c(\mu),\mu)$ is a function of the hard scales, EM current conservation is enforced by the transverse projector, and the ellipses represent subleading terms. At leading power, the nonlocal and local FFs are simply related by $W_\gamma(-Q^2) \simeq c_3\, f_+(-Q^2)/2$ where $f_+$ is the local $K \to \pi$ transition FF. Since the latter scales as $f_+ \simeq 16\pi \alpha_s f_K f_\pi/Q^2$ due to hard gluon exchange and $c_{3} \sim m_c^2/(16\pi^2 Q^2)$ from GIM cancellation in the limit $m_c=0$, we have $W_\gamma(-Q^2) \sim (\alpha_s/\pi) \Lambda^2 m_c^2/Q^4$. The four-quark operators $\sim (\bar{s}_L \gamma_\alpha u_L)(\bar{u}_L \gamma_\beta d_L)$ included in the ellipsis in Eq.~\eqref{eq:ope} couple to the valence quark in $K^+=s\bar{u}$ (see Fig.~\ref{fig:figure1d}). These `weak annihilation' (WA) effects are probably the dominant source of power corrections since their matching coefficients are tree level. The WA contribution to $W_\gamma$ scales as $\sim 16\pi \alpha_s f_K f_\pi/Q^2 \times (\Lambda^3/Q^3)$, suppressed by $\sim 16\pi^2 \Lambda^3 /(m_c^2 Q)$ compared to the leading term in the OPE. The power corrections can be made arbitrarily small by taking $Q \gg m_c$, `unraveling' the resummation accomplished in Ref.~\cite{Buras:2006gb}, at the expense of spoiling the convergence of the dispersive integral Eq.~\eqref{eq:emFFdisp}. These scaling arguments will be solidified by the result of an explicit OPE matching calculation.

\section[Correspondence]{The correspondence between $\bm{K^+ \to \pi^+ \ell^+ \ell^-}$ and $\bm{K^+ \to \pi^+ \nu \bar{\nu}}$}
Both vector and axial currents contribute to the nonlocal HME in $K^+ \to \pi^+ \nu \bar{\nu}$ which defines vector, axial, scalar and pseudoscalar FFs as follows,
\begin{align}
    & \frac{-i}{q^2} \int d^4x e^{iqx} \braket{\pi^+(p)|T \{J_{V(A)}^\mu(x)\hat{H}(0)\}|K^+(p+q)} \nonumber \\
    &\quad \quad \quad \quad= W_{V(A)}(q^2) \left[(2p+q)^\mu - q^\mu \frac{M_K^2-M_\pi^2}{q^2}\right] + W_{S(P)}(q^2) \left[q^\mu \frac{M_K^2-M_\pi^2}{q^2} \right] \, . \label{eq:FZ}
\end{align}
Here the vector current is $J_V^\mu = \sum_q g_V^q \bar{q}\gamma_\mu q$, where $g^{u,c}_V=\frac{1}{2}-\frac{4}{3}\sin^2 \theta_W$, $g^{d,s}_V=-\frac{1}{2}+\frac{2}{3}\sin^2\theta_W$ and the axial current is $J_A^\mu = \sum_q g_A^q \bar{q}\gamma_\mu \gamma^5 q$, where $g_A^{u,c}=\frac{1}{2}$, $g_A^{d,s}=-\frac{1}{2}$. The scalar and pseudoscalar FFs do not contribute to the $K^+ \to \pi^+ \nu \bar{\nu}$ amplitude since $m_\nu=0$. We can establish a correspondence between the time ordered products in Eqs.~\eqref{eq:Fgamma} and~\eqref{eq:FZ} using the following exact relations,
\begin{align}
    J_V&= \frac{g_V^u}{e_u}J_\gamma - \frac{1}{4}(J_d+J_s) = \frac{g_V^u-g_V^d}{e_u-e_d} J_\gamma - \frac{1}{6}(J_u+J_d+J_s+J_c) \, . \label{eq:currentrelation}
\end{align}
It is straightforward to verify that the equations above are exact by inserting the definitions of the currents $J_q^\mu = \bar{q}\gamma^\mu q$ and comparing the coefficients of each flavor. 

In the first relation in Eq.~\eqref{eq:currentrelation}, the connected $u,c$ contributions to $J_V$ have been absored into $J_\gamma$. The OPE relates the two nonlocal FFs to each other for large $Q$ when the disconnected terms are perturbatively suppressed, as $W_{V}(-Q^2) = (g_V^u/e_u) W_\gamma(-Q^2) + O(\Lambda/Q)$.
The QCD corrections to $W_{V,\gamma}$ at leading power (and the nontrivial $m_c$ dependence therein) are universal up to normalization and have been absorbed in the relation above. The universality is only spoiled at leading power by disconnected three-loop corrections in which the currents couple to $d,s$ quarks. Additionally, one can show that $W_{A}(-Q^2) = -(g_A^u/e_u)W_\gamma(-Q^2) + O(m_c/Q)$, where (for $m_c/Q\ll 1$) the sign of the axial current coefficient appears since left-handed quarks propagate in the loops and $P_L \gamma_5 =-P_L$. The charm mass dependence of this relation can be taken into account, although as far as we are aware, this dependence has not been calculated at NLO in QCD.

In the second relation in Eq.~\eqref{eq:currentrelation}, the isovector contribution to $J_V$ has been absorbed into $J_\gamma$. On the $\pi^+ \pi^-$ branch cut, only the isovector current contributes, and only to the vector FF, so $\mathrm{disc}\,W_V(q^2) = (g_V^u-g_V^d)/(e_u-e_d) \,\textrm{disc} W_\gamma(q^2)$ and $\mathrm{disc}\, W_A(q^2) = 0$.
The relations above are exact in the isospin limit and below inelastic thresholds $q^2<(3M_\pi)^2$. Above the three-pion cut the isoscalar component of the electromagnetic current also contributes, both to $W_V$ and $W_A$. 

Finally, inserting the definitions of the electroweak couplings in terms of the Weinberg angle, we obtain the summary relations between $W_{V-A}=W_V-W_A$ and $W_\gamma$, 
\begin{align}
W_{V-A}(-Q^2)& \simeq \left( \frac{3}{2}-2\sin^2 \theta_W\right) W_\gamma(-Q^2) \, , & (Q \gg \Lambda,m_c) \label{eq:relation1} \\
\textrm{disc}\,W_{V-A}(q^2) & \simeq \left( 1 - 2 \sin^2 \theta_W \right) \,\textrm{disc}\,W_\gamma(q^2) \, & (q^2 \lesssim 1\,\textrm{GeV}^2) \label{eq:relation2}
\end{align}
The two conditions taken together, linked by dispersive representations of \emph{both} form factors should provide a strong constraint on $W_{V-A}(q^2)$ for the neutrino mode, given the magnitude and phase of the electromagnetic form factor $W_\gamma(q^2)$. The magnitude and phase of the latter should be obtained from a joint analysis of $K^+ \to \pi^+ \ell^+ \ell^-$ and $K \to 3\pi$, respectively, outlined in the following section.

\section{Hadronic amplitudes from pion rescattering}
\label{sec:KT}

The problem of predicting the $K^+ \to \pi^+ \ell^+ \ell^-$ distribution and/or calculating the corresponding nonlocal matrix element in $K^+ \to \pi^+ \nu \bar{\nu}$ reduces to determining the $K^+ \pi^- \to \pi^+ \pi^-$ partial wave amplitude which enters in the discontinuity of the electromagnetic FF in Eq.~\eqref{eq:discFgamma}. The remaining task is to determine such hadronic amplitude as accurately as possible, consistent with the $K^+ \to \pi^+ \pi^+ \pi^-$ Dalitz distribution. However, in turn one immediately runs into the problem of extracting the phase of the $P$-wave projection of the $K^+\pi^- \to \pi^+ \pi^-$ amplitude.

To obtain this phase, following a recent development in the field, we consider a phenomenological, yet rather model-independent approach, based on isospin symmetry and truncation of the partial wave expansion. In Ref.~\cite{Bernard:2024ioq} it was shown for the first time that there is in principle sufficient information from an exhaustive isospin analysis of all four dominant CP-conserving $K \to 3\pi$ decays to constrain the magnitude and phases of all nine (four $\Delta I = 1/2$ + five $\Delta I = 3/2$) reduced isospin amplitudes in the decomposition of a generic $K \to 3\pi$ amplitude of the form
\begin{align}
    \braket{\pi_b \pi_c|\hat{H}_{\Delta I}|K_i \pi_a}&= \sum_{I,\mathcal{I}}C_{iabc}^{\Delta I,I,\mathcal{I}}\braket{I||\hat{H}_{\Delta I}||\mathcal{I}} \, ,\label{eq:cgcoeffs}
\end{align}
where $I=0,1,2$ and $\mathcal{I}=1/2, 3/2$ are the isospins of the $\pi\pi$ and $K\pi$ states, respectively. The coefficients are known products of Clebsch-Gordon coefficients. The reduced matrix elements $\mathcal{A}_{\Delta I}^{I,\mathcal{I}}(s,t)=\braket{I||\hat{H}_{\Delta I}||\mathcal{I}}$ which pass the selection rules $|\mathcal{I}-\Delta I|\leq I\leq\mathcal{I}+\Delta I$ are functions of $s=(p_K-p_a)^2=(p_b+p_c)^2$ and $t=(p_K-p_b)^2$.

For our application it is appropriate to consider a less general description of $K \to 3\pi$ amplitudes, focusing mainly on the charged kaon decays of interest. This simplification is plausible since the $\Delta I = 1/2$ and $\Delta I =3/2$ reduced amplitudes for the $K^+$ decays can be written in the form
\begin{align}
    \mathcal{A}(K^+ \to \pi^+\pi^+\pi^-) &= \mathcal{F}(t,s,u) + \mathcal{F}(u,t,s) + \mathcal{H}(s,t,u) \, , \\
    \mathcal{A}(K^+ \to \pi^+ \pi^0 \pi^0) &= \mathcal{F}(s,t,u) + \mathcal{H}(s,t,u) \, ,
\end{align}
where $\mathcal{F}(s,t,u)$ and $\mathcal{H}(s,t,u)$ are certain linear combinations of reduced matrix elements. Crossing symmetries enforce a specific form of these in terms of single-variable functions $F_I$ and $H_I$,
\begin{align}
    \mathcal{F}(s,t,u) &= F_0(s) + (s-u)F_1(t) + (s-t)F_1(u) - \frac{2}{3}F_2(s) + F_2(t)+F_2(u) \, , \\
    \mathcal{H}(s,t,u) &= \frac{3}{2}(s-u)H_1(t) + \frac{3}{2}(s-t)H_1(u) + H_2(s) - \frac{1}{2} H_1(t) - \frac{1}{2}H_1(u) \, .
\end{align}
Analyticity and unitarity is enforced by the following dispersive representations of each single variable function, where the discontinuities are given recursively by partial wave unitarity
\begin{align}
    F_I(s) &= P_{n-1}^{[F_I]}(s;\alpha_i)+\frac{s^n}{\pi}\int ds' \, \frac{ \textrm{disc} \, F_I(s')}{{s'}^n(s'-s)} \, \\
    \textrm{disc} \, F_I(s)&= \sin \delta_\ell^I(s) e^{-i\delta_I^\ell(s)} \left[F_I(s)+ \sum_{J} \int_{-1}^1 dz \, K_{IJ}(s,t(s,z))  F_{J}(t(s,z)) \, \right] \label{eq:Fhat}
\end{align}
and similarly for $H_I(s)$. The polynomials $P_{n-1}$ of order $n-1$ correspond to the low-energy expansions of the single-variable functions in terms of $n$ subtraction parameters $\alpha_i$.

The cross-channel rescattering kernel $K$ is a known function involving the coefficients in Eq.~\eqref{eq:cgcoeffs} and kinematical quantities. In the absence of the second term in the discontinuities, corresponding to left-hand cuts of the single-variable functions, the KT equations reduce to a set of decoupled, linear homogeneous integral equations, with the analytical solutions given by
\begin{align}
F_I(s)=\Omega_\ell^I(s)P_{n-1}^{[F_I]}(s;\alpha_i) \, , \quad \Omega_\ell^I(s) = \exp \left[ \frac{s}{\pi}\int ds' \, \frac{\sin \delta_\ell^I(s')}{s'(s'-s)} \right] \, .
\end{align}
The Omnes functions $\Omega_I^\ell$ (with $\ell=0$ for $I=0,2$ and $\ell=1$ for $I=1$) unitarize the single-variable functions with respect to direct-channel pion rescattering, and depend on the $\pi \pi$ phase shifts $\delta_0^0$, $\delta_1^1$ and $\delta_0^2$ taken as external input, for instance, from pion-nucleon cross-sections and the pion vector form factor. We emphasize that we have not yet considered $K\pi$ phase shifts in our analysis, although such effects may be needed for precision phenomenology.

\begin{figure}[ht]
\centering
\begin{minipage}{\textwidth}
\includegraphics[width=0.24\textwidth]{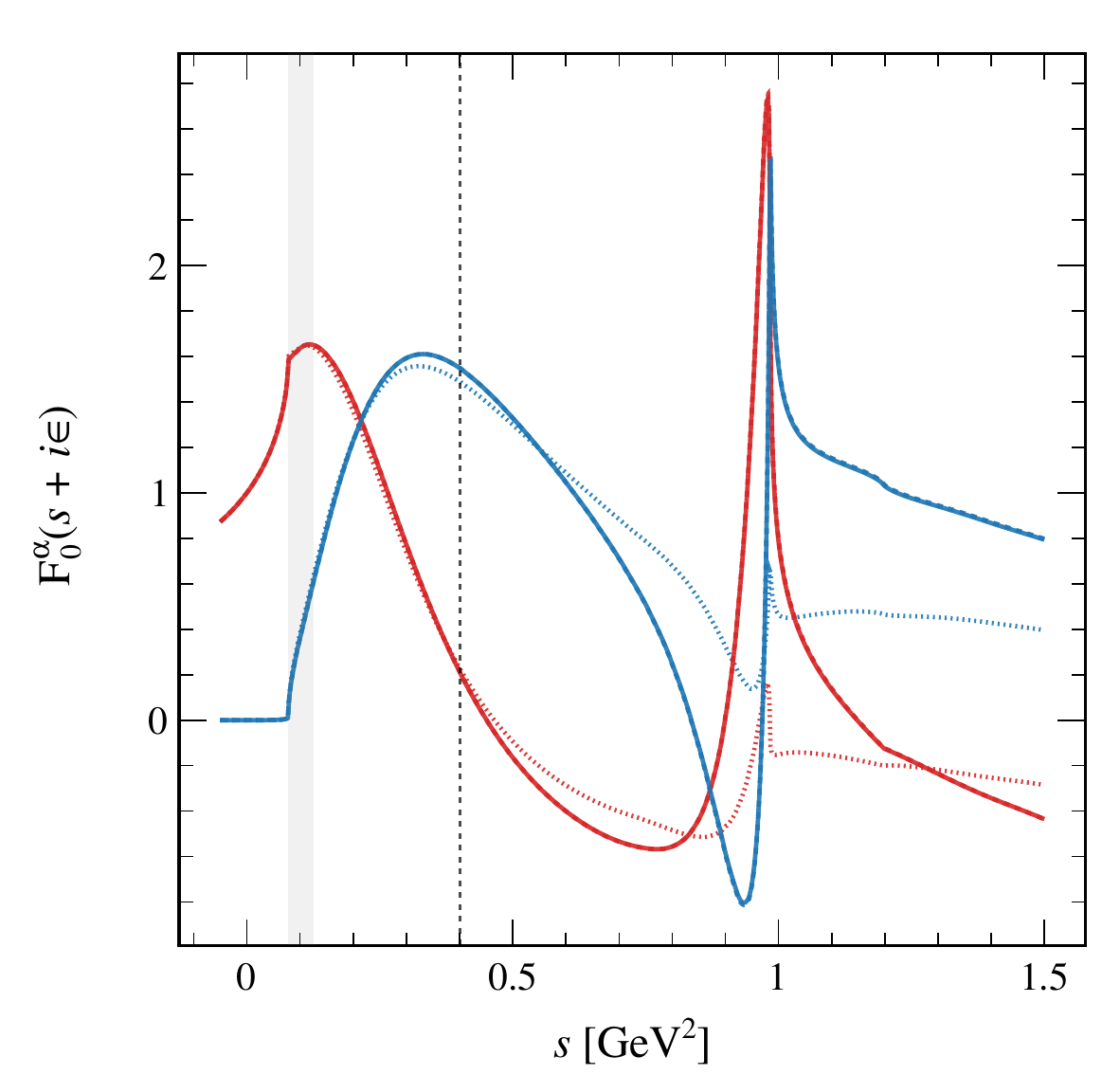}
\includegraphics[width=0.24\textwidth]{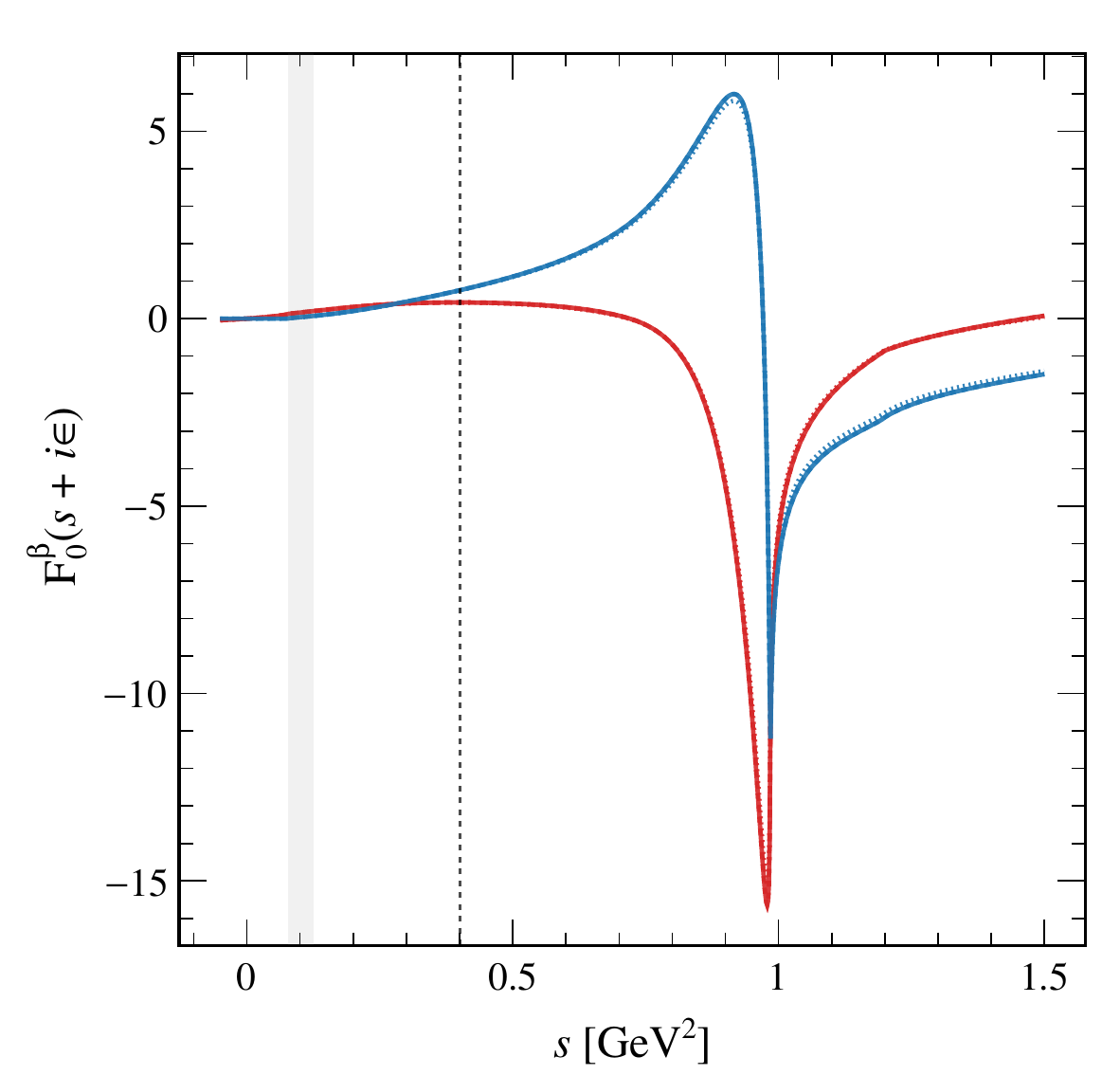}
\includegraphics[width=0.24\textwidth]{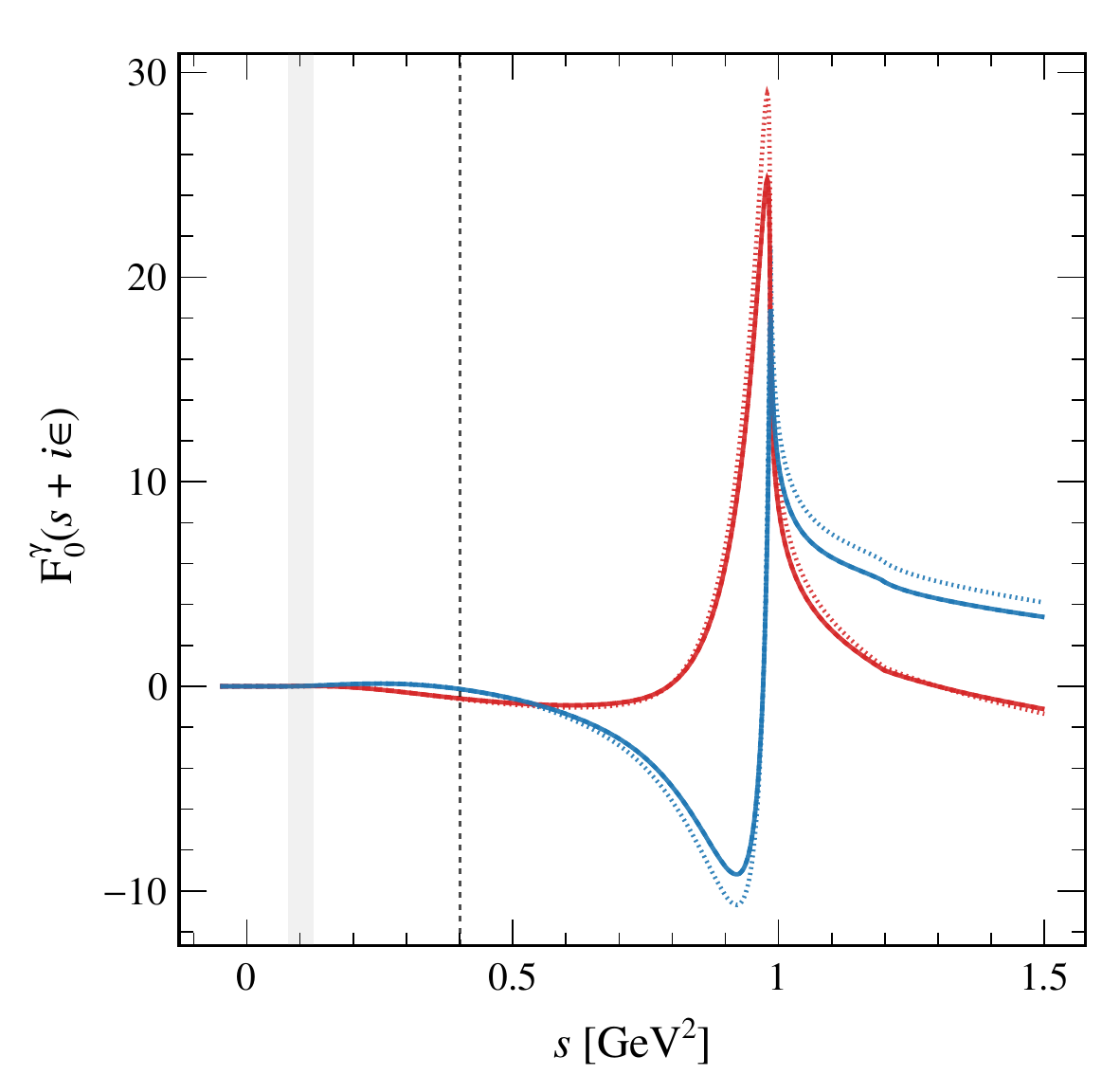} \\
\includegraphics[width=0.24\textwidth]{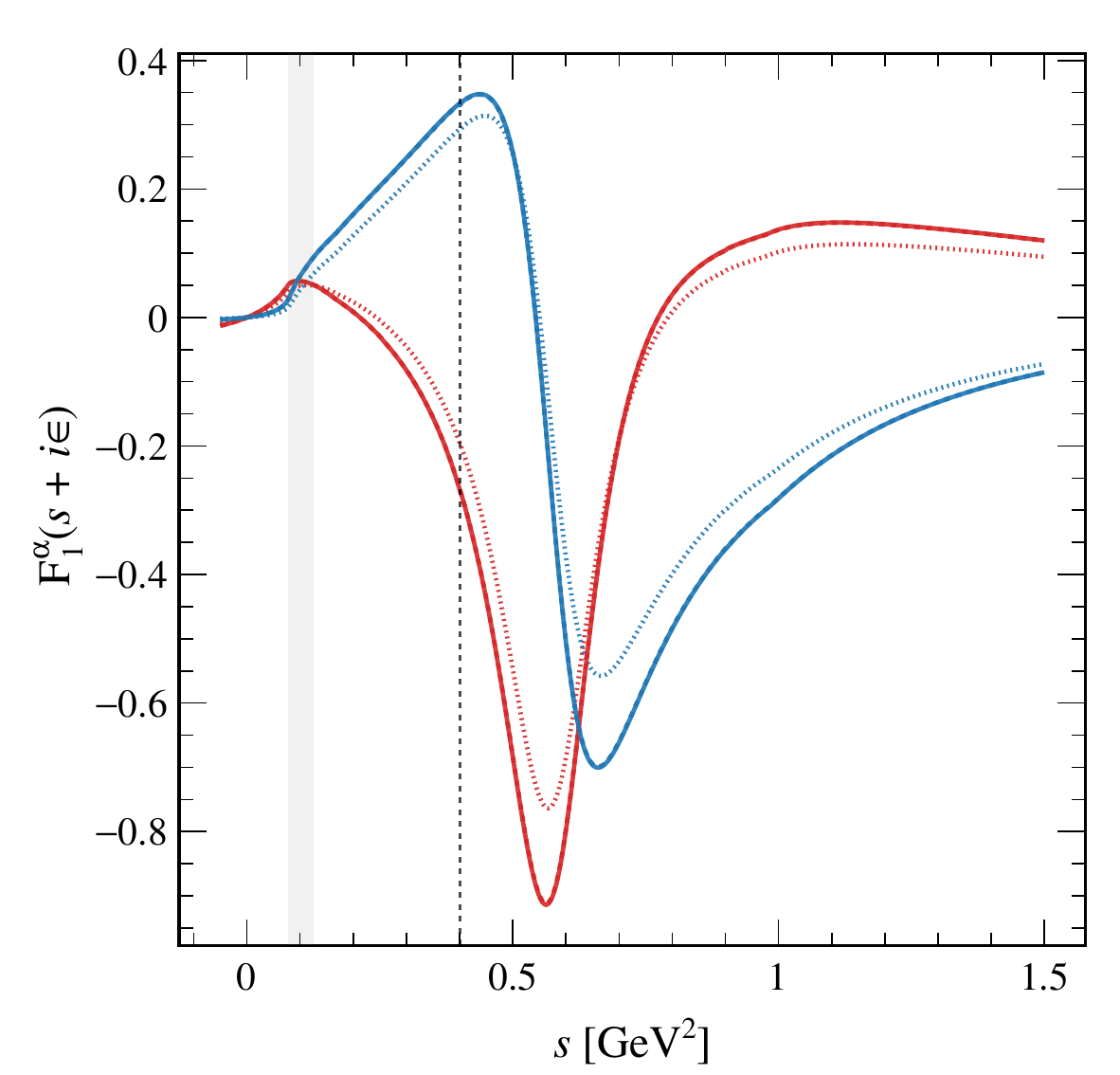}
\includegraphics[width=0.24\textwidth]{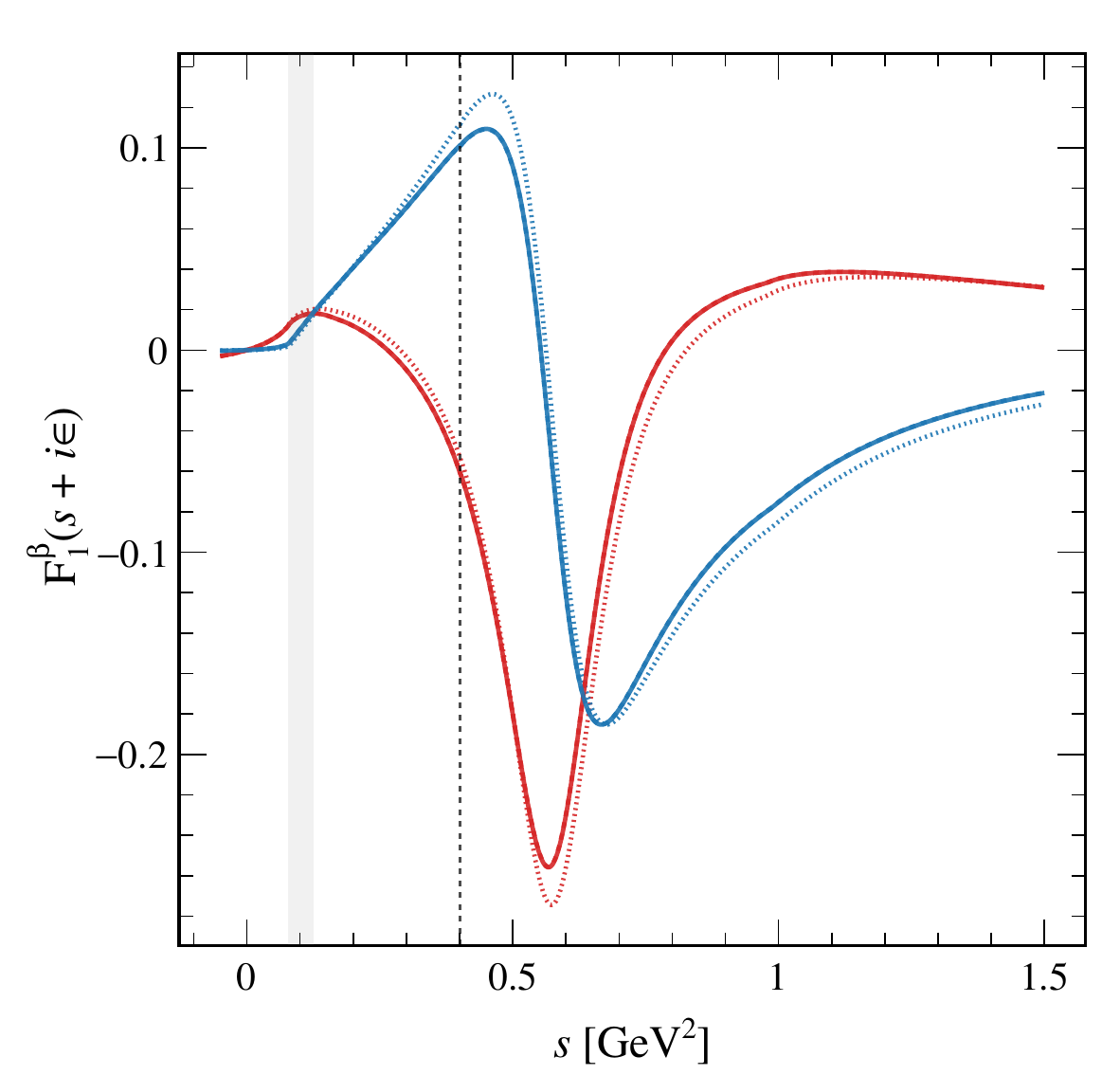}
\includegraphics[width=0.24\textwidth]{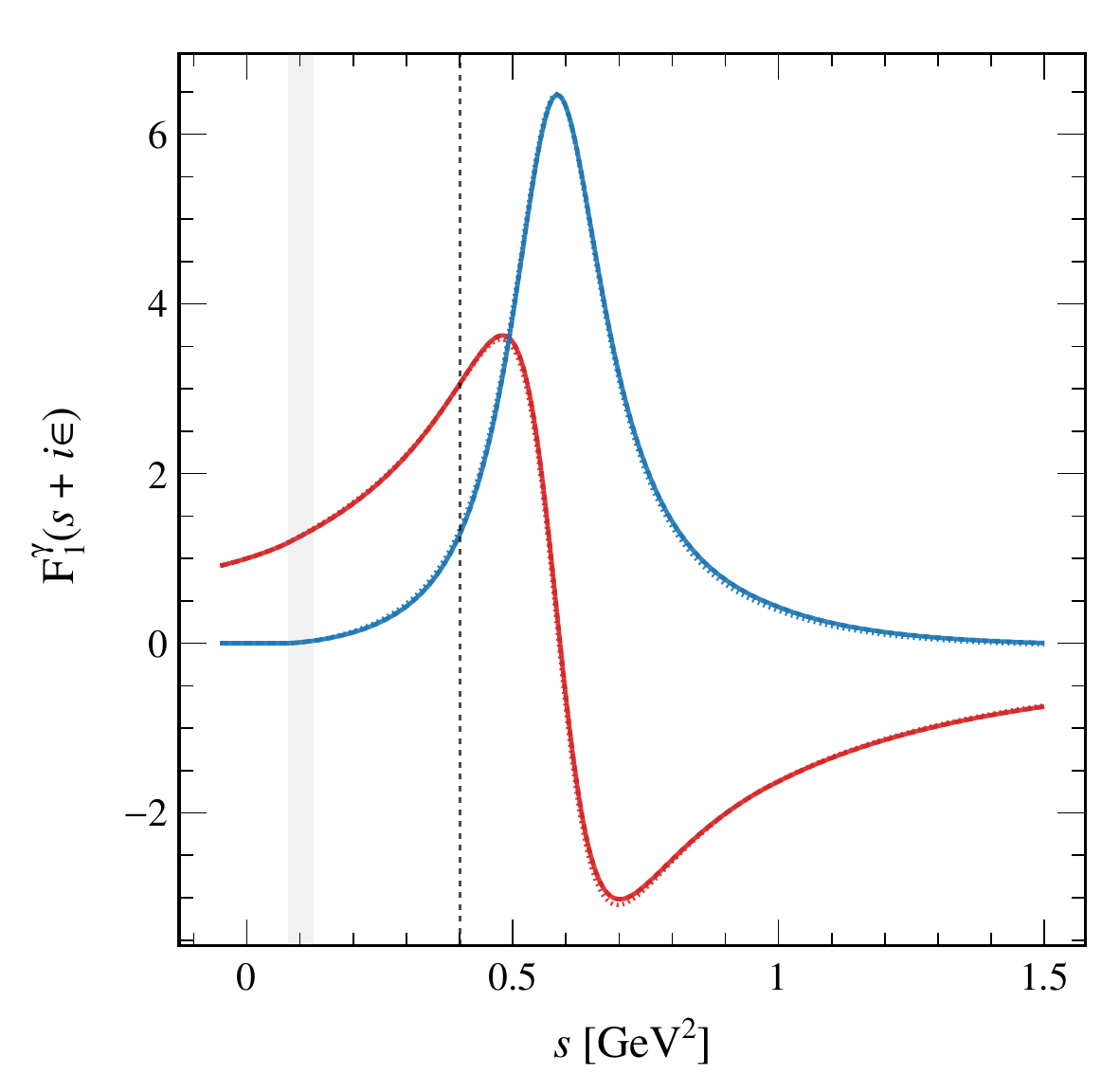} 
\includegraphics[width=0.24\textwidth]{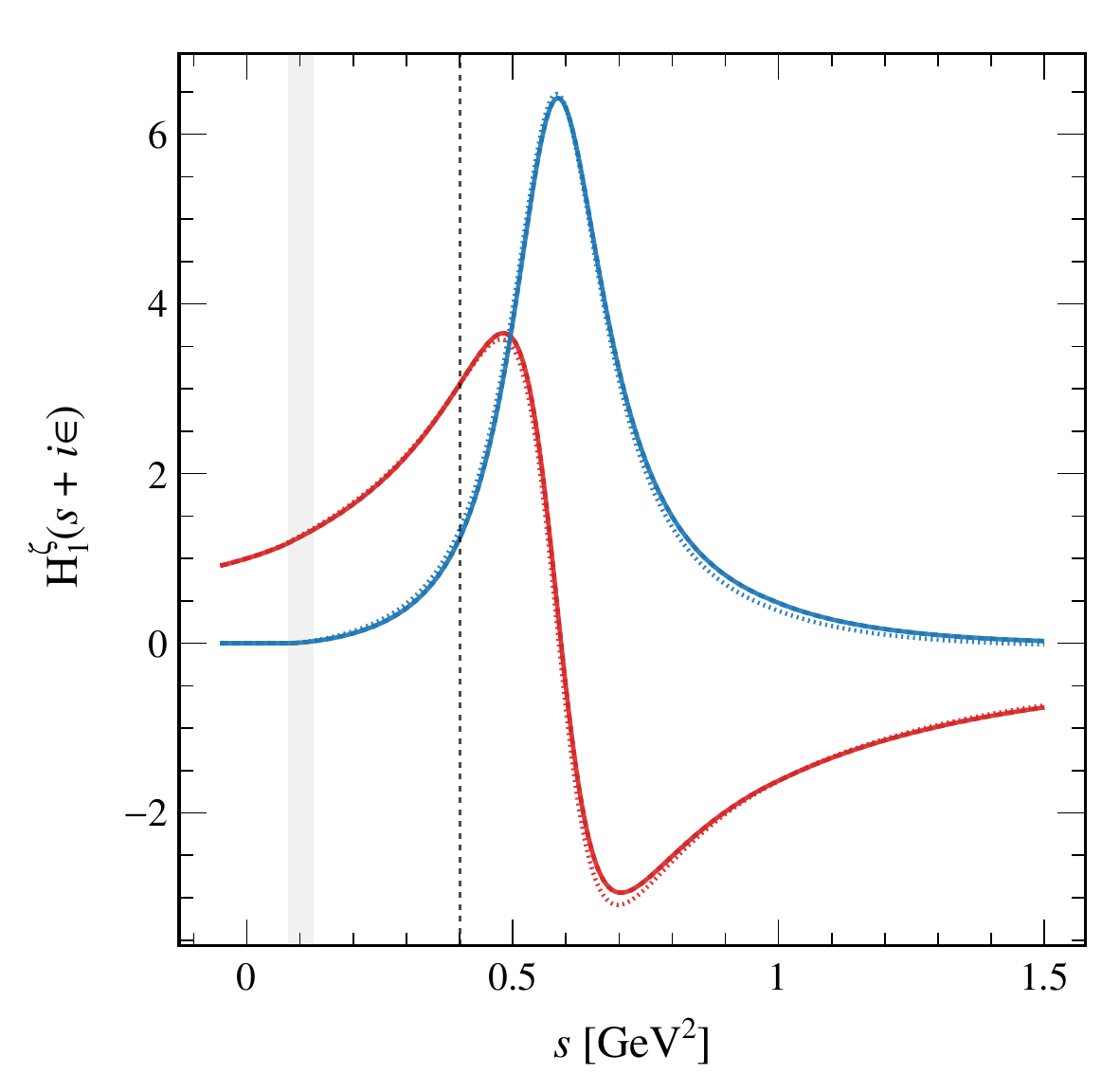}
\\
\includegraphics[width=0.24\textwidth]{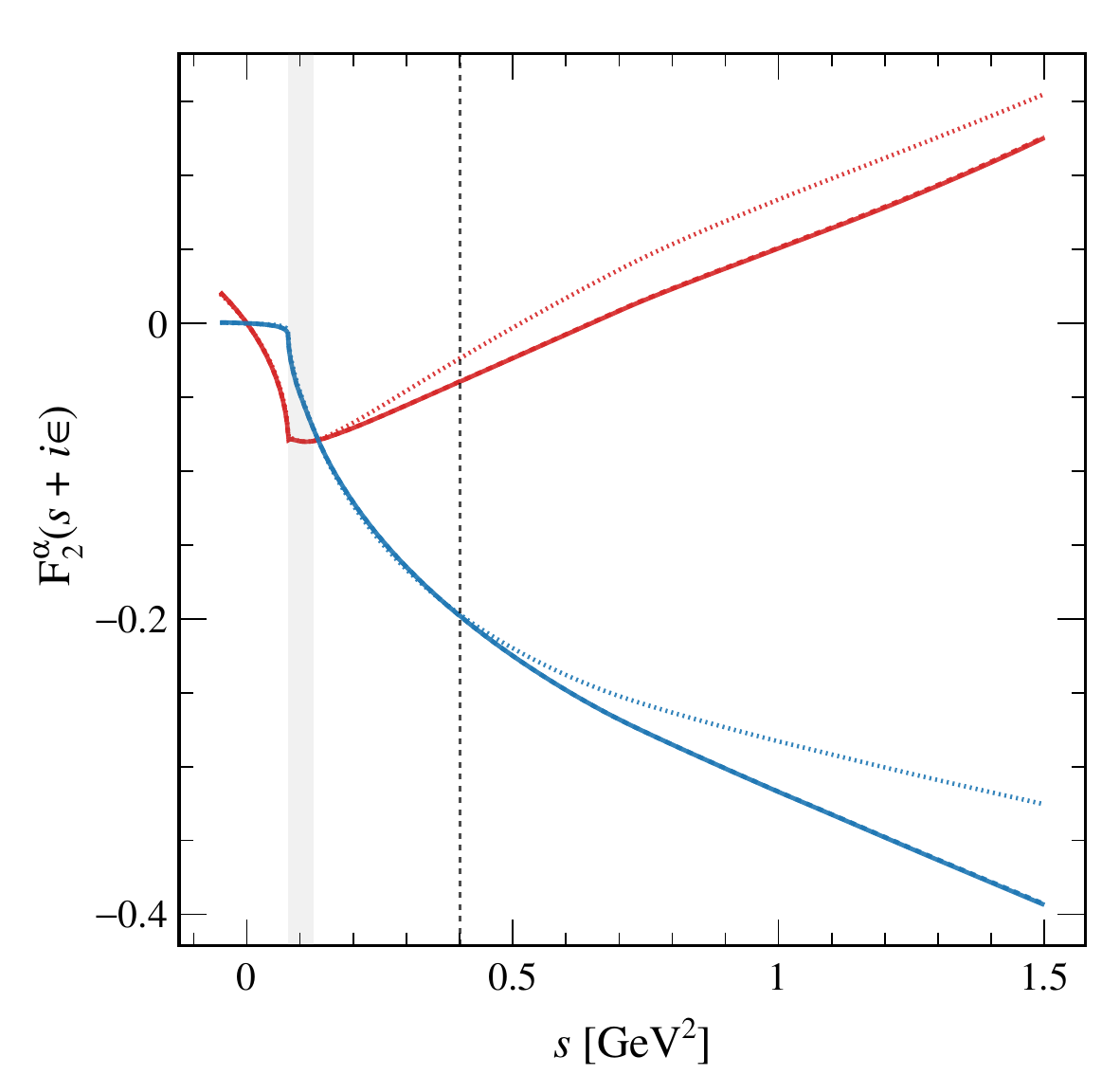}
\includegraphics[width=0.24\textwidth]{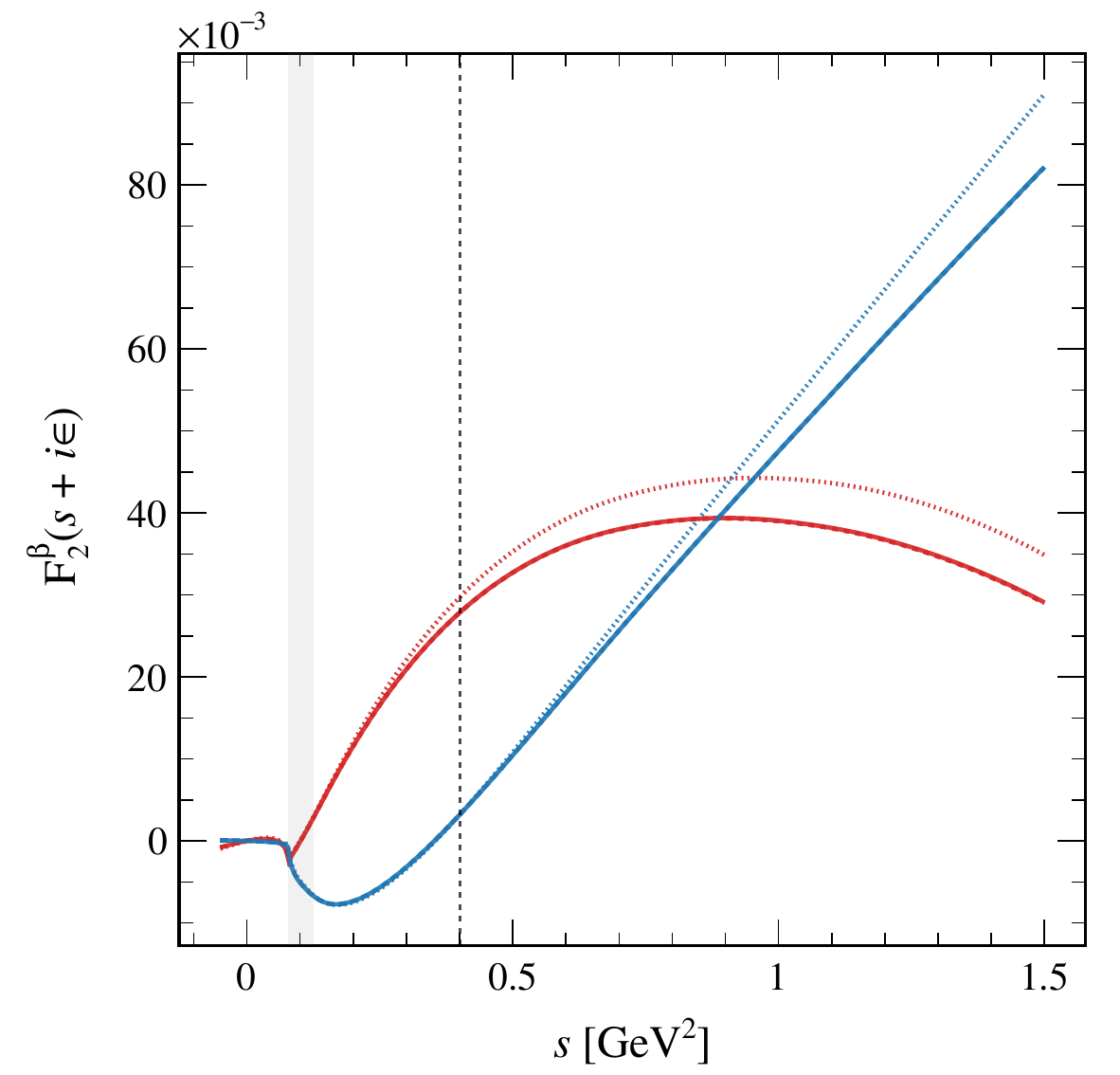}
\includegraphics[width=0.24\textwidth]{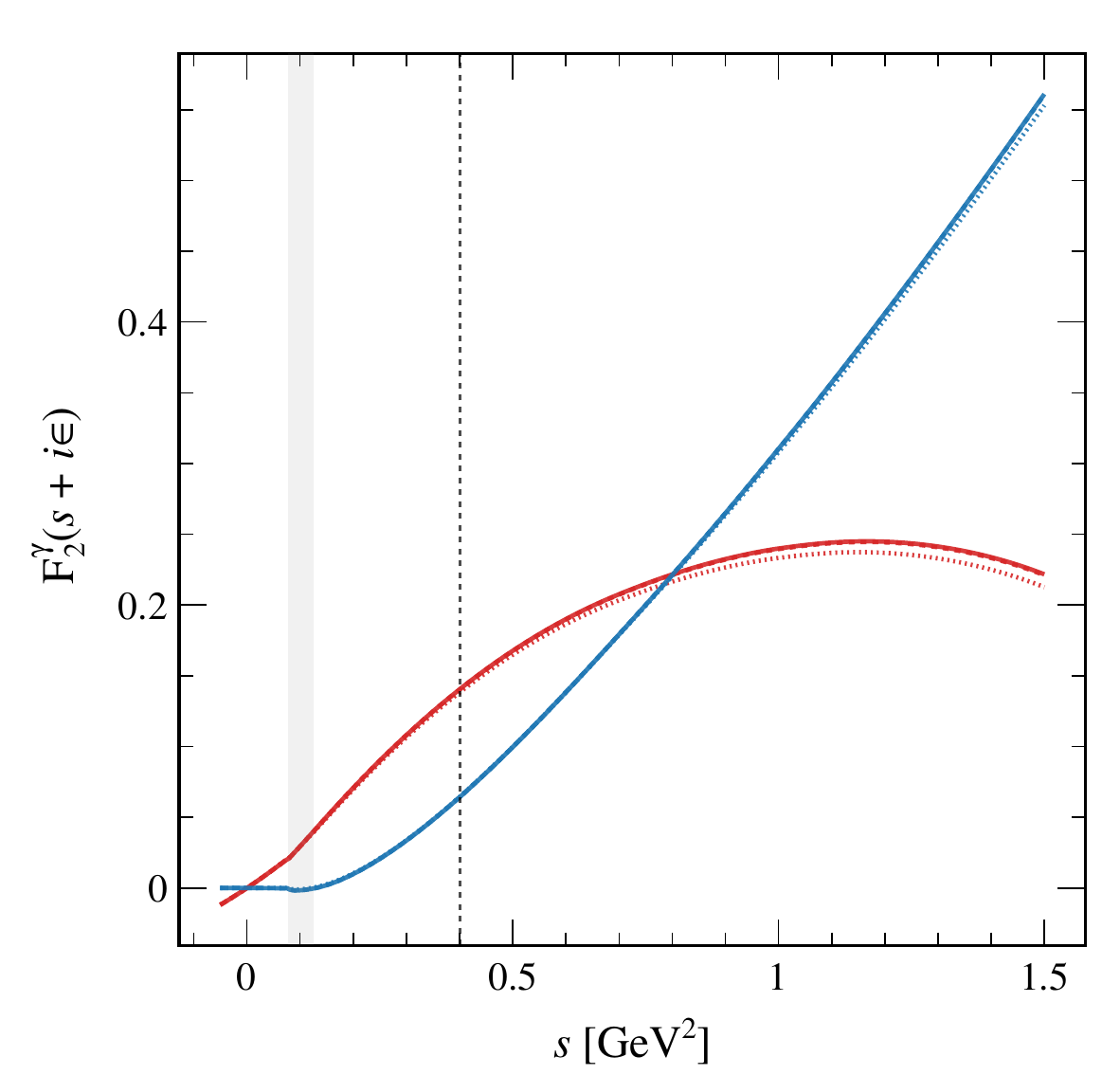}
\includegraphics[width=0.24\textwidth]{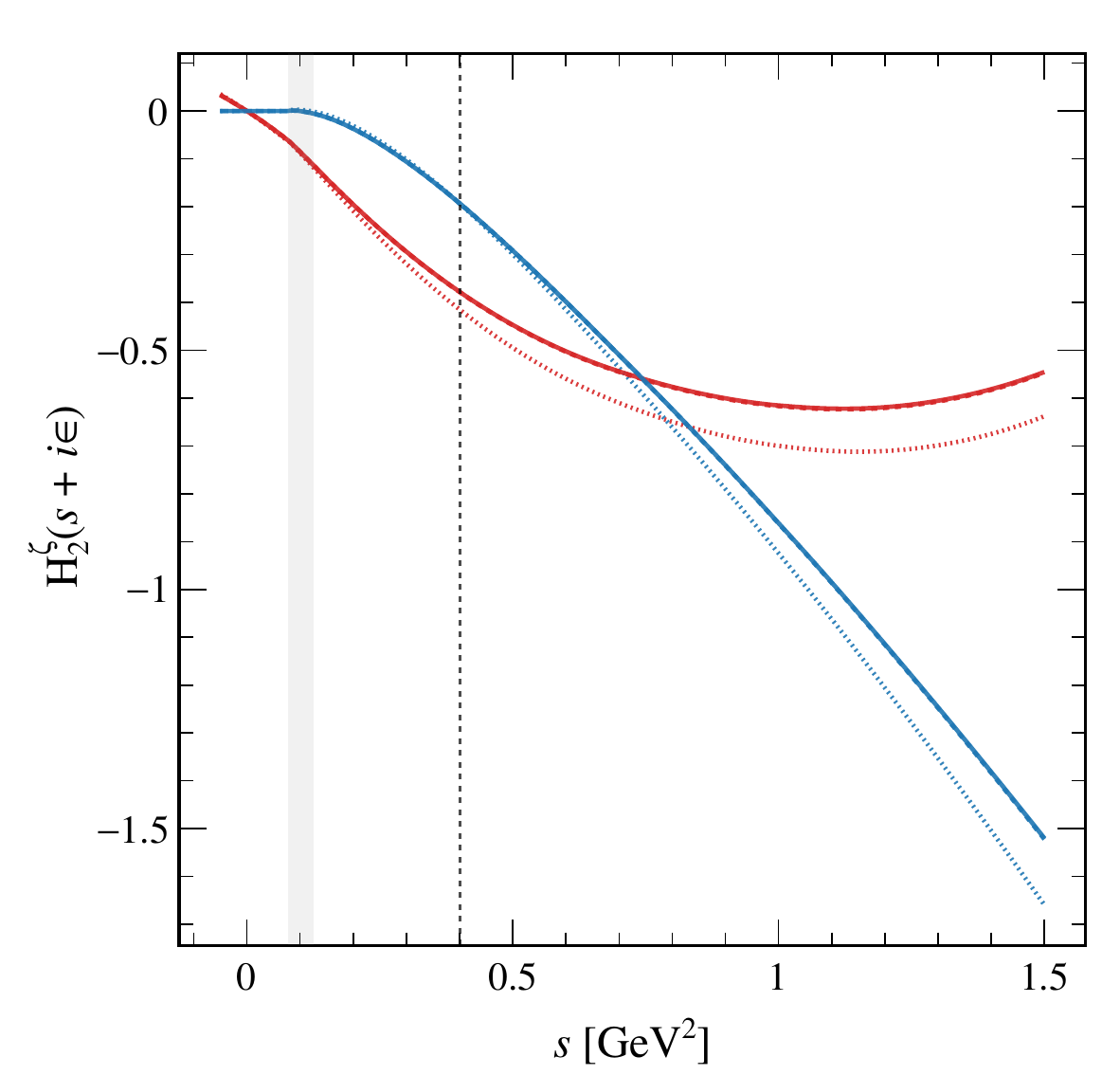}
    \caption{Expanded view of the real (red) and imaginary (blue) parts of all isobars after one (dotted) three (dashed) and six (solid) iterations of the KT equations. The projections of the $K \to 3\pi$ Dalitz region are indicated by the thin gray bands. Here we have taken $M_K=M_{K^+}$ and $M_\pi=M_{\pi^+}$. \label{fig:out}}
\end{minipage}
\end{figure}

Specific assumptions of the high-energy behavior of each amplitude lead to limits on how many subtractions (and therefore degrees-of-freedom) are allowed. We consider more constraining assumptions as Ref.~\cite{Bernard:2024ioq} and introduce only four subtractions in the form $P[F_0]=\alpha+\beta s$, $ P[F_1]=\gamma$, $P[H_1]=\zeta$ (with $P[F_2] = P[H_2]=0$). The isospin amplitudes can thus be expressed in the form
\begin{align}
F_I(s) &= \Omega_\ell^I(s)\left[ \alpha F_I^\alpha(s) + \beta F_I^\beta(s) + \gamma F_I^\gamma(s) \right] \, , \quad H_I(s) = \Omega_\ell^I(s) \left[ \zeta H_I^\zeta(s) \right]
\end{align}
where the so-called isobars $F_I^{\alpha_i}(s)$ are independent of the values of the subtractions $\alpha_i$. The isobars are obtained as solutions to the coupled, inhomogenous linear system of KT equations. We solve these equations in the standard way by iterating them with initial conditions specified by the Omnes solutions. The comprehensive result of this exercise is shown in Figure~\ref{fig:out}. Finally, what remains is to fit the four subtractions to the eight observables in $K^+ \to \pi^+ \pi^+ \pi^-$ and $K^+ \to \pi^0 \pi^0 \pi^+$: the two branching fractions, two linear slope parameters and four quadratic slope parameters, or an appropriate subset thereof, for instance a subset less sensitive to isospin breaking effects.

\section{Summary}

We have studied the implications of dispersive representations of the nonlocal form factors in semileptonic $K^+ \to \pi^+ \ell^+ \ell^-$ and $K^+ \to \pi^+ \nu \bar{\nu}$ decays, respectively $W_\gamma$, $W_{V(A)}$. All three form factors $W_i(-Q^2)$ can be identified with one another up to normalization for asymptotic $Q \gg \Lambda$ via their connection to the local $K \to \pi$ form factor in the OPE. Additionally the discontinuities of $W_{V,\gamma}$ in the physical region are precisely related in the isospin limit via the universality of the pion vector form factor. Here we propose a data-driven method in which the $K^+ \to \pi^+ \ell^+ \ell^-$ distribution along with an isospin amplitude analysis of $K \to 3\pi$ decays can be leveraged to obtain the discontinuity of the electromagnetic form factor above $\pi^+ \pi^-$ threshold and below inelastic thresholds, anchoring the discontinuity of $W_{V-A}$ in the same domain using Eq.~\eqref{eq:relation2}. 

While this framework is based on symmetries and perturbative techniques grounded in the OPE, we have found that isospin breaking corrections have an important systematic effect on intermediate results for the $K \to 3\pi$ amplitude analysis. A detailed assessment of these sources of theoretical uncertainty is necessary to interpret any results obtained under the assumption of isospin symmetry. Finally, we should point out that a complete description of nonlocal amplitudes in $K^+ \to \pi^+ \nu \bar{\nu}$ should include $WW$ exchanges (Figs.~\ref{fig:figure1c} and~\ref{fig:figure1e}) which cannot be straightforwardly parameterized in terms of HMEs depending only on a single variable $q^2$.

\bibliographystyle{JHEP}
\bibliography{biblio_proceeding}

@article{NA62:2022qes,
    author = "Cortina Gil, Eduardo and others",
    collaboration = "NA62",
    title = "{A measurement of the $K^+\to \pi^+ \mu^+\mu^-$ decay}",
    eprint = "2209.05076",
    archivePrefix = "arXiv",
    primaryClass = "hep-ex",
    reportNumber = "CERN-EP-2022-189",
    doi = "10.1007/JHEP06(2023)040",
    journal = "JHEP",
    volume = "11",
    pages = "011",
    year = "2022",
    note = "[Addendum: JHEP 06, 040 (2023)]"
}

@article{NA482:2009pfe,
    author = "Batley, J. R. and others",
    collaboration = "NA48/2",
    title = "{Precise measurement of the $K^\pm \to \pi^\pm e^+e^-$ decay}",
    eprint = "0903.3130",
    archivePrefix = "arXiv",
    primaryClass = "hep-ex",
    reportNumber = "SLAC-PUB-14772, CERN-PH-EP-2009-005",
    doi = "10.1016/j.physletb.2009.05.040",
    journal = "Phys. Lett. B",
    volume = "677",
    pages = "246--254",
    year = "2009"
}

@article{NA62:2024pjp,
    author = "Cortina Gil, Eduardo and others",
    collaboration = "NA62",
    title = "{Observation of the $K^{+}\rightarrow\pi^{+}\nu\bar{\nu}$ decay and measurement of its branching ratio}",
    eprint = "2412.12015",
    archivePrefix = "arXiv",
    primaryClass = "hep-ex",
    reportNumber = "CERN-EP-2024-343",
    month = "12",
    year = "2024"
}

@article{Isidori:2005xm,
    author = "Isidori, Gino and Mescia, Federico and Smith, Christopher",
    title = "{Light-quark loops in $K \to \pi \nu \bar{\nu}$}",
    eprint = "hep-ph/0503107",
    archivePrefix = "arXiv",
    reportNumber = "RM3-TH-05-2",
    doi = "10.1016/j.nuclphysb.2005.04.008",
    journal = "Nucl. Phys. B",
    volume = "718",
    pages = "319--338",
    year = "2005"
}

@article{Buras:2006gb,
    author = "Buras, Andrzej J. and Gorbahn, Martin and Haisch, Ulrich and Nierste, Ulrich",
    title = "{Charm quark contribution to $K^+ \to \pi^+ \nu \bar{\nu}$ at next-to-next-to-leading order}",
    eprint = "hep-ph/0603079",
    archivePrefix = "arXiv",
    reportNumber = "TUM-HEP-600-05, IPPP-05-74, DCPT-05-148, TTP06-06, ZU-TH-23-05, FERMILAB-PUB-05-512-T",
    doi = "10.1007/JHEP11(2012)167",
    journal = "JHEP",
    volume = "11",
    pages = "002",
    year = "2006",
    note = "[Erratum: JHEP 11, 167 (2012)]"
}

@article{Falk:2000nm,
    author = "Falk, Adam F. and Lewandowski, Adam and Petrov, Alexey A.",
    title = "{Effects from the charm scale in $K^+ \to \pi^+ \nu \bar{\nu}$}",
    eprint = "hep-ph/0012099",
    archivePrefix = "arXiv",
    reportNumber = "CLNS-00-1707, JHEU-TIPAC-20007",
    doi = "10.1016/S0370-2693(01)00343-4",
    journal = "Phys. Lett. B",
    volume = "505",
    pages = "107--112",
    year = "2001"
}

@article{Mescia:2007kn,
    author = "Mescia, Federico and Smith, Christopher",
    title = "{Improved estimates of rare K decay matrix-elements from $K_{\ell 3}$ decays}",
    eprint = "0705.2025",
    archivePrefix = "arXiv",
    primaryClass = "hep-ph",
    doi = "10.1103/PhysRevD.76.034017",
    journal = "Phys. Rev. D",
    volume = "76",
    pages = "034017",
    year = "2007"
}

@article{Buchalla:1998ba,
    author = "Buchalla, Gerhard and Buras, Andrzej J.",
    title = "{The rare decays $K\to \pi \nu\bar\nu$, $B \to X \nu\bar\nu$ and $B \to l^+ l^-$: An Update}",
    eprint = "hep-ph/9901288",
    archivePrefix = "arXiv",
    reportNumber = "CERN-TH-98-369, TUM-T31-337-98",
    doi = "10.1016/S0550-3213(99)00149-2",
    journal = "Nucl. Phys. B",
    volume = "548",
    pages = "309--327",
    year = "1999"
}

@article{Lunghi:2024sjy,
    author = "Lunghi, E. and Soni, A.",
    title = "{Light quark loops in $K^\pm \to \pi^\pm \nu\bar\nu$ from vector meson dominance and update on the Kaon Unitarity Triangle}",
    eprint = "2408.11190",
    archivePrefix = "arXiv",
    primaryClass = "hep-ph",
    month = "8",
    year = "2024"
}

@article{Bernard:2024ioq,
    author = "Bernard, V\'eronique and Descotes-Genon, S\'ebastien and Knecht, Marc and Moussallam, Bachir",
    title = "{A dispersive study of final-state interactions in $K\rightarrow \pi \pi \pi $ amplitudes}",
    eprint = "2403.17570",
    archivePrefix = "arXiv",
    primaryClass = "hep-ph",
    doi = "10.1140/epjc/s10052-024-13084-y",
    journal = "Eur. Phys. J. C",
    volume = "84",
    number = "7",
    pages = "744",
    year = "2024"
}

@article{Misiak:1999yg,
    author = "Misiak, Mikolaj and Urban, Jorg",
    title = "{QCD corrections to FCNC decays mediated by Z penguins and W boxes}",
    eprint = "hep-ph/9901278",
    archivePrefix = "arXiv",
    reportNumber = "TUM-HEP-336-98, IFT-14-98",
    doi = "10.1016/S0370-2693(99)00150-1",
    journal = "Phys. Lett. B",
    volume = "451",
    pages = "161--169",
    year = "1999"
}

@article{Bernard:2025xyn,
    author = "Bernard, V{\'e}ronique and Descotes-Genon, S{\'e}bastien and Knecht, Marc and Moussallam, Bachir",
    title = "{A dispersive approach to the CP conserving $K\to\pi\ell^+\ell^-$ radiative decays}",
    eprint = "2510.17316",
    archivePrefix = "arXiv",
    primaryClass = "hep-ph",
    month = "10",
    year = "2025"
}

@article{DAmbrosio:1998gur,
    author = "D'Ambrosio, G. and Ecker, G. and Isidori, G. and Portoles, J.",
    title = "{The Decays $K \to \pi \ell^+ \ell^-$ beyond leading order in the chiral expansion}",
    eprint = "hep-ph/9808289",
    archivePrefix = "arXiv",
    reportNumber = "INFN-NA-IV-98-25, UWTHPH-1998-45, FTUV-98-61, IFIC-98-62",
    doi = "10.1088/1126-6708/1998/08/004",
    journal = "JHEP",
    volume = "08",
    pages = "004",
    year = "1998"
}

\end{document}